# Collection security management at university libraries: assessment of its implementation status

**A. A. Maidabino** [1] and **A.N. Zainab** [2]
[1] Department of Library & Information Science, Faculty of Education,
Bayero University Kano, NIGERIA
[2] Faculty of Computer Science and Information Technology,
University Malaya, Kuala Lumpur, MALAYSIA
e-mail: maidabinoum@yahoo.com; zainab@um.edu.my

**ABSTRACT**

*This study examines the literature on library security and collection security to identify factors to be considered to develop a collection security management assessment instrument for university libraries. A "house" model was proposed consisting of five factors; collection security governance, operations and processes, people issues, physical and technical issues and the security culture in libraries. An assessment instrument listing items covering the five factors was pilot tested on 61 samples comprising chief librarians, deputy librarians, departmental, sectional heads and professional staff working in four university libraries in Nigeria. The level of security implementation is assessed on a scale of 1=not-implemented, 2=planning stage, 3=partial implementation, 4=close to completion, and 5=full implementation. The instrument was also tested for reliability. Reliability tests indicate that all five factors are reliable with Cronbach's alpha values between 0.7 and 0.9, indicating that the instrument can be used for wider distribution to explore and assess the level of collection security implementation in university libraries from a holistic perspective.*

## INTRODUCTION

University libraries face a number of security challenges with their collections (both print and non-print). Library collections constitute the bedrock for services provided to the community and serve as important assets to the library. As such, securing and protecting the collections can help libraries provide an effective service in response to the information needs of the university community. Collection security implies the need for libraries to provide, maintain and secure its collection to ensure longetivity, accessibility and effective provision of services to users. To achieve this noble objective however, libraries need an effective strategy to assess the degree of collection security, breaches they are facing and establish an acceptable level of collection security implementation. This study is an attempt to provide an assessment instrument that can be used by university library managers to assess collection security implementation in their libraries.

## OBJECTIVES

The objectives of this paper are to:

a) Identify the factors that should be considered when assessing the status of collection security implementation in university libraries;
b) Map the factors and the items into a collection security management assessment instrument; and
c) Test the reliability and validity of the collection security assessment instrument.

## LITERATURE REVIEW

The aim of university libraries is to provide access to information resources in both print and non print formats. Balancing access and security in libraries is a difficult but a necessary task. A number of studies have described how crimes and security breaches incidences can affect the provision of library services to users. Lorenzen (1996) and Holt (2007) identified several such incidents, (i) theft of physical materials; (ii) theft or alteration of data; and (iii) theft of money as major security crime in libraries. Other forms of breaches include non-return of items by borrowers, theft of library equipment, personal theft (from staff and users), verbal and physical abuse against staff and users, and vandalism against library buildings, equipment and stock destruction, all of which can directly or indirectly affects the provision of library services (Ewing 1994). Similarly, Lorenzen (1996) reported how different forms of collection mutilation such as underlining and highlighting text in library books, tearing and or removing pages of books and annotating in books margins can temper with the subject-content of library collection, thereby making it unusable to users. Wu and Liu (2001) identified the aim of a modern university library as largely to provide access to both print and non-print collections and this makes it necessary to develop a balance between ownership and access to information or knowledge. This can be achieved by proper planning strategy including the planning for access control in line with the security requirement and the present and future mission or goals of the parent institutions. Ajegbomogun (2004) identified the types of security breaches in university libraries, which included theft and book mutilation and reasoned the cause to security lapses, insufficient or limited number of essential materials, and user's financial constraints. Ameen and Haider (2007) opined that access to collection is important as this service has supported scholarship in the humanities, sciences and social sciences and remains the key to intellectual freedom. Similarly, university libraries need to create an environment where primary resource materials are respected, handled carefully, and returned intact to the collection so that they might be studied again in the future. Therefore, materials that are not meant to be used by patrons should not be accessible to them. For example, the unprocessed materials should be kept in a secured area; public access to special and rare collections should be monitored and physically protected to prevent vandalism, theft and other security breaches (Rude and Hauptman 1993). Studies conducted by Ajegbomogun (2007), Bello (1998), and Holt (2007) identified rare books, manuscripts and special collections as frequent target of theft and mutilation because of the special demand for in depth studies of such materials. The above studies indicate that the processes that handle access to collection such as acquisition, technical processes, circulation, shelving and storage of items in libraries need to be considered from the security perspectives and assessed by a collection security measurement instrument.

Most of the published literature on library security issues focuses on specific types of security breach. Theft, mutilation and vandalism are highly covered by research articles. Boss (1984) highlighted theft and arson as threats to collections and proposed that libraries formulate a planned security measure to protect their collections. Boss also identified physical weaknesses in libraries in terms of unsecured windows, faulty emergency exits, unstaffed computer rooms, poor policies and procedures, lack of security plans, poor security points (exits, loading areas, windows, special collections) inadequate loans and renewal periods, lack of security manuals and poor signage as some of the causes of security breaches. Ewing (1994) identified abuses in UK libraries, which included book and non-book theft, non-return of borrowed items, verbal and physical abuse, and vandalism against library buildings and properties. Ewing also reported an estimated collection lost rate of 2.6% and that is between 1500 and 3000 books stolen annually. Abifarin (1997), Allen (1997) and Bello (1998) reported high rate of book theft, mutilation and misplacing of books in Nigerian academic libraries. They suggested measures to reduce the problems, which include tightening security at library entrances and exits, expulsion of students involved in theft and mutilation, provision of multiple copies of heavily used text, reducing the cost of photocopying, and periodic searching of students hostels and staff offices. Atkins and Weible (2003) believe that successful inventorying process helps identify missing items; however it may be dependent on the size of the library's collection. They proposed using interlibrary loan (ILL) data failure cases to identify materials missing from a library's collection instead. Brown and Patkus (2007) stressed that university libraries must ensure that access and storage areas for collection are arranged and monitored for quick and easy inspection. Special and rare collections in particular need to be stored separately, with separate folders within the collection so that they can be easily checked by the staff. Furthermore, a reliable and effective procedure for accessibility to such collection must to be created. Accessibility to library collections can also be enhanced by proper supervision and control of the library environment, especially designated areas for library assets. University library management must ensure that access to any area within the library is clearly defined and regulated. Staff should also enforce restrictions by challenging, in a non-confrontational manner, any unauthorised user found to be outside the designated public areas (Houlgate and Chaney 1992) These studies highlight the importance of considering the security aspects of physical and infrastructural perspective of library buildings and facilities to ensure collection security, thus implicating a factor that needs to be included in the assessment instrument.

Another factor is the human aspect of library security. This involves creating the right atmosphere for greater security awareness amongst library staff, users and the university community at large. Omoniyi (2001) found that both students and staff were often involved in collection theft and this may be due of their unawareness of the graveness of the thieving issue. Holt (2007) highlighted theft of library collections by staff as a real problem that libraries should address and not ignore because of the risk of bad publicity. Holt suggested several methods to deal with staff theft including the installation of high security lock systems, tightening of collection transportation and movement procedures, marking collections to indicate ownership, good record keeping and undertaking periodic inventories. He also highlighted the need for libraries to cultivate professional culture and behaviour with regard to safeguarding the library's collection and the need for library management to take the lead in developing an honest culture with reporting responsibility.

Lowry and Goetsch (2001) highlighted the importance of creating shared culture of mutual responsibility for security and safety of library collections. This involves making clear to users and staff about the safety and security policies and guidelines in libraries, especially those regarding food consumption in the library, theft, mutilation, and disruptive behaviour. They also emphasised on policies regarding training of staff to create an awareness culture. The people aspect of library security issues, include staff's nonchalant attitudes to users' needs and ignorance about security issues (Ives, 1996). There are instances which indicate that library staff entrusted to protect the integrity, accessibility and confidentiality of library materials were often the source of collection security problems (Ajegbomogun 2004). These studies indicate that the people aspect and security culture are important factors when assessing collection security in libraries.

Another factor covered by the literature is the importance of good governance. Brown and Patkus (2007) proposed a security plan that comprises these components: a written security policy; the appointment of a security manager; a security survey conducted to assess current and projected needs; identifying preventive measures; ensuring a secure premise for both during and after working hours; installation of a security system; ensuring collection security through regular inventory; proper storage area; marking collections to establish ownership and instituting a tracking system of lost and over borrowed items; and managing, educating and training users and staff. The Association of College and Research Libraries (2006) proposed a guideline for the security of rare books, manuscripts and special collections. The guidelines proposed the establishment of proper governance by hiring library security officers who plan and administer security programmes, prepare and spearhead written policies. The library is also advised to closely monitor the entrances and exits of special collection reading areas, making staff aware of collection security problems, providing training in security measures, monitoring users in the stacks, reading and reference areas, keeping adequate accession records, and aiding access through proper cataloguing records and finding aids. The importance of good and supportive governance with clear policies and procedures in order to maintain an acceptable level of collection security in libraries is therefore necessary.

Other studies focus on security breaches like purposive misshelving of items, especially reference books (Alao et al. 2007), disruptive behaviour as a result of drunkenness and drug addiction (Lorenzen 1996; Ardndt 1997; Momodu 2002; Ajegbomogun 2004), natural and man-made disaster (Evans et al. 1999; Shuman 1999; Aziagba and Edet 2008) and demand outstripping supply, which may give rise to delinquent behaviour such as stealing, mutilating or using another user's borrowing tickets (Bello 1998). All of which may subsequently remain a serious threat to the security of the library and its collection.

From the published literature, the researchers have identified the factors that should be considered when assessing collection security implementations in libraries. These factors were identified, derived and used in this study to formulate the library collection security assessment instrument which is framed on a "house" model described in the following section.

## THE COLLECTION SECURITY MANAGEMENT MODEL (CSMM)

The factors that comprise collection security management in libraries are derived from published literature. The factors are then positioned in a "house" for collection security management model (CSMM) for libraries. The house adopts and adapts the operational model proposed by Da Veiga and Eloff (2007), who have used a house to frame the information system security governance. The CSMM likened library collection security governance to a secured house where the alarm system installed should provide adequate protection. Nevertheless, even in this secure situation, security may be breached if the owner leaves the house with the front door unlocked. This illustrates that security measures would be ineffective if the behaviour of those in the home or an organisation are nonchalant about implementing the security processes. The model can be viewed from five factors; governance, processes, people, physical as well as technology perspectives and the existence of security culture in libraries. The framework should provide university and library management with a working instrument to assess and implement a more holistic approach to collection security management.

The term collection security in this context refers to protecting collection from unauthorised use, displacement, defacement, modification and destruction. Libraries protect the following attributes of their collections (Corporate Governance 2004).

- Confidentiality – This infers that the collection is available only to those authorised at the various levels (controlled access to types of registered members).
- Integrity – The library has to make sure that the collection or the information they carry is not altered, accurate and complete. Therefore, management of its security needs to insulate the collection from accidental or deliberate change to the contents. Accuracy refers to proper description of collections, which are shelved or stored appropriately. Completeness refer to collections that are not mutilated, missing, decayed, miss placed, over borrowed, insulated from deliberate or accidental change, and secure from theft or vandalism.
- Availability – This refer to making sure that authorised users have reliable and timely access to collection at the time they need it (timeliness) and at the promised times (appropriate opening hours) and through a reliable network system, which makes items available without delay.

The model describes a holistic plan for collection protection in libraries, combining the governance, process, people, physical and cultural factors to ensure that a reasonable level of collection security management is in place, hence minimizing risks to the library's main assets, its collections. In the model all factors are of equal weight in ensuring that the confidentiality, integrity and availability of the library's collections are maintained at a reasonable level. The factors in the house model are detailed in Figure 1 and will be used as the basis to design the assessment instrument.

## The Governance

Governance in this context refers to the provision of a set of roles, policies and responsibilities and practices exercised by members of a security team responsible for formulating objectives and policies, ensuring that objectives and policies are achieved, ascertaining that risks are identified and managed appropriately (Omoniyi 2001; Allen and Westby 2007). The governance factor stresses the importance of considering collection security as part of management responsibility (Brown and Patkus 2007). This means that, university libraries should design a concise strategy with plans that articulate the vision and direction for risk management (Purtell 2007). The library security management (LSM) team ideally should be chaired by senior personnel with representation from various departments, and members from the university's security department. The LSM team members should have the necessary experience and knowledge about collection security issues and management so that they can command authority to manage and ensure collection security compliance. Risk assessment is undertaken to identify risk at various levels within the library setting so that security planning is prioritized (Cowan 2003). Regular risk assessment of collection involves compiling detailed information about library assets in terms of types and value of collections, identifying threats, vulnerabilities, possible risk situations, and estimating the cost of loss (cost analysis). This would show serious commitment to identify collection security problems and implement programmes to mitigate it. Collection security breaches can be identified through meetings, questionnaire, observations, and reports. This should be followed by identifying what should be done to reduce risk, monitoring, revising and evaluating the risk process. The Governance factor stresses the need for library and collection managers to document, maintain, review and update risk policies and procedures as well as prepare reports via newsletters, interactive web pages or other in house publications to publicize collection security initiatives and create awareness amongst employees and users (Saffady 2005). This factor provides evidence whether good collection security governance is in place in libraries.

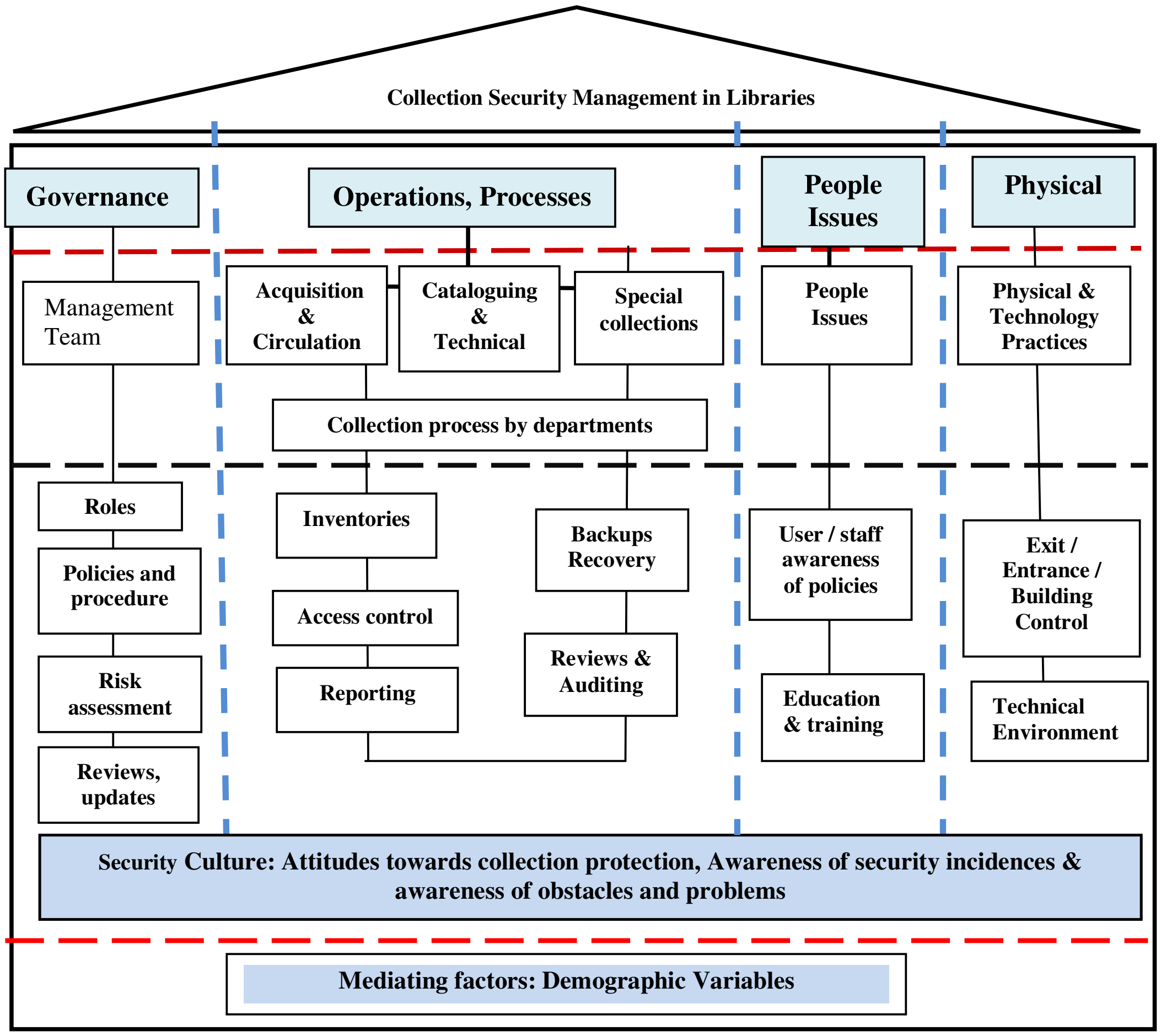

Figure 1: The House Model for Collection Security Management for Libraries (CSML).

## The Operational Processes

This factor involves the processes of putting into operation security programmes formulated by the security management team through the relevant departments. These are; (a) the acquisition department, which is involved in accessioning and marking items to establish ownership, maintaining an inventory list which can be used to identify missing or misplaced or cost of items, and to facilitate backups and recovery processes (Holt 2007; Brown and Patkus 2007; Association of College and Research Libraries 2006; Atkins and Wible 2003; Omoniyi 2001); (b) the circulation department, which shelves and stores items for quick and easy inspection by users, creates manual or computer systems to record and track the use of the collections, control access, undertaking stock taking and inventory, report on delinquent borrowings, items that are lost, misplaced, stolen, abused, decayed or damaged (Luurtsema 1997); (c) the cataloguing and technical department, which processes and documents collections the library's catalogue system, verifies ownership of circulated items as well as attach identification marks to establish ownership and ensures that the status of unprocessed items are reported and access to them are controlled (Brown and Patkus 2007; Omoniyi 2001); (d) the special collections, which involves preserving and conserving collections, controlling and monitoring access, proper inspection of the collection before and after use and providing insurance coverage for valuable collections (Swartzburg, Bussey and Garretson 1991).

## The People Factor: Training Programmes

This factor involves the human or people aspects of the model, particularly programmes involving staff being trained, retrained and made aware of policies and procedures on collection security management processes. This should include security awareness being formalised in organisational policy and procedures and communicated to every employee who works with information resources (Saffady 2005). It stipulates the need to determine collection security roles and responsibilities in university libraries and ways to handle, supervise and monitor qualified and trained staff. Staff's knowledge of the availability of training programmes will help them handle security incidences, prepare reliable and useful reports.

## Physical and Technology Factors

This factor involves both the physical and technical mechanisms in implementing a secure collection environment. The physical environment refers to the safety and security of the library premises which holds the collections (Swartzburg, Bussey and Garretson 1991). The physical security measures should begin with the physical architecture of the building or management of space where collections are held, controlling building entrances and exits; requiring identifications to access to general as well as rare and special collections areas; and scheduling patrols within building parameters. The technical aspect consists of the technology practices and procedures that the collection security programmes embraced. It refers to electronic security system and devices to handle collection security processes, control security breaches, and installation at strategic entry points of the library. This includes security systems such as electronic anti-theft devices, visual cameras, smoke detection and alarm system at entrances, exits and stack areas in the library. This system will help prevent unauthorised removal of collections and feasible monitoring and detection of user traffic in general reading and reference rooms, as well as shelves areas (Ameen and Haider 2007; Omoniyi 2001).

## Security Culture

This factor encompasses acceptable user and staffs' attitudes and awareness toward the importance of protecting collections in the library. Awareness is an unseen element but is demonstrated through perceptions; such as (a) staff's attitudes about the importance of security policies and processes; (b) their awareness of security breaches; and (c) the limitations of implementations (Lowry and Goetsch 2001; Holt, 2007). Lowry and Goetsch (2001) refer to this situation as shared culture of mutual responsibility for security and safety where staffs are provided with the information and the tools to respond to a variety of situations and are able to take action when called upon to do so. This attitude and awareness "glues" or "grounds" the effectiveness of security governance, management and operations. This importance is based on the premise that "organisations do not change, but people do, and therefore people change organisations" (Verton 2000).

## The Demographics

The demographics of the respondents and their libraries are also provided in the house model as moderating factor. This part of the house model includes socio-demographics of the respondents such as gender, educational qualification, job title, years of working experience, and occupational rank. While the information about the libraries focused on size of the library collections, size of the total staff and size of the professional staff respectively. In the context of this study, the demographics are features of house model which are expected to support the implementation of collection security management.

## METHODOLOGY

In this study the context chosen are Nigerian university libraries. This study reports on the first phase of an exploratory approach which describes the extraction of factors that encompasses collection security governance in published literature to formulate an assessment instrument, which is subsequently piloted and tested for reliability. The second phase of the research (not part of this paper) involves using the instrument in a wider context. Due to time and cost constraint the instrument was piloted at four universities which are conveniently chosen because they are close to Bayero University (where the researcher works at). In total, 87 questionnaires were distributed between November 2008 and January 2009 to university librarians, chief librarians, deputy librarians, departmental and sectional heads and professional staff working in the selected four university libraries. A total of 61 (70.1%) returned questionnaires were found usable. All the libraries chosen for the pilot study had been established for more than 10 years and with different collection sizes. It is therefore assumed that the libraries have adopted some degree of collection security governance. The demographics of respondents are shown in Tables 1 and 2.

Table 1: Information about the Libraries

| The university samples | Year of establishment | Staff population | Size of the collections |
|---|---|---|---|
| Usman Danfodiyo University, Sokoto | 1970-1979 | 0-49 | 251,000-500,00 |
| Bayero University Kano | 1970-1979 | 100-149 | 501,000-750,000 |
| Ahmadu Bello University | 1960-1969 | 150-199 | 751,000-1000,000 |
| University of Maiduguri | 1970-1979 | 100-149 | 251,000-750,000 |

Table 2: Response Rate based on Gender and by Institutional Distribution

| The university sampled | Male | Female | Total | Percentage |
|---|---|---|---|---|
| Usman Danfodiyo University Sokoto | 6 | 2 | 8 | 13% |
| Bayero University Kano | 8 | 4 | 12 | 20% |
| Ahmadu Bello University | 14 | 5 | 19 | 31% |
| University of Maiduguri | 15 | 7 | 22 | 61% |
| Total | 43 | 18 | 61 | 100% |
| Percentage | 70% | 30% | 100% | 100% |

**The Collection Security Management Assessment Instrument**

The collection security management assessment instrument is designed to help university's library management to determine the status of their collection security management implementation. The instrument is intended to serve as a bench mark to highlight the factors that nee consideration when ensuring the security of collections in university libraries.

The instrument consists of six sections in line with the CSMM (Figure 1). The first section contained 14 items that assess the governance aspect of collection security in university libraries. The second section covers the general processes aspect of collection security such as acquisition and circulation, technical cataloguing, and special collection. The third section focuses on the human or staff training issues. The fourth covers the physical and technological factors and the fifth section focuses on security culture which encompasses awareness of collection incidences, attitudes towards collection protection, and awareness of obstacles and problems for collection protection. All the five components of this tool are equally depicted in the House Model. The last part of the instrument solicits demographic details about the university libraries and the respondents. A summary of the components and their items of the instrument is provided in Table 3. The factors laid out in the model together with the items covering it are rated on a five point Likert-type scale (1= Strongly Disagree, 2 = Disagree, 3 = Fairly agree, 4 =Agree, 5= Strongly Agree).

Table 3: The Factors with the Number of Items of the Instrument

| Sections and Factors Covered | Number of items |
|---|---|
| Section 1: Governance of collection security | 14 |
| Section 2: General operations and processes of collection security implementation | 29 |
| Section 3: People Issues in collection security management | 10 |
| Section 4: Physical factors (non-electronic and electronic) | 12 |
| Section 5: Collection security culture | 43 |
| Section 6: Demographic information about respondents and libraries | 11 |

The Keiser-Meyer-Olkin (KMO) measure of sampling adequacy was examined for all the items. Kaiser (1974) reported in Field (2005) recommended accepting values greater than 0.5 as acceptable KMO values, as values below this would require researchers to either collect more data or rethink which variables to include. Furthermore, Field indicates that, values between 0.5 and 0.7 are mediocre, 0.7 to 0.8 are good, 0.8 to 0.9 as great and values above 0.9 as superb. In this study the KMO values for all the items are greater than 0.5 which indicates that the data is considered acceptable for factor analysis.

## LEVEL OF COLLECTION SECURITY IMPLEMENTATION IN LIBRARIES

The level of implementation of all factors can be determined by totaling the scores of all the items listed under each factor. The minimum and maximum scores for each factor are dependent on the number of items listed under each section. Table 4 indicates the measuring scale used for all factors, where, the minimum will be 14 and the maximum varies from 20 to 105. Under each factor the performance level is measured in accordance to the following scale: 1 = Non-implementation, 2 = Planning stage, 3 = Partial implementation, 4 = Close to completion and 5 = Full Implementation.

Table 4 provides the mean scores for all the factors of collection security management of the instrument by university rating. The mean score of each university is compared with a scale provided together with the mean score which will indicate the level of collection security management implementation.

Table 4: Rating Scale Used to Assess the Level of Collection Security Implementation

| | | | **Mean score by University Libraries** | | | | **Implementation levels** | | | | |
|---|---|---|---|---|---|---|---|---|---|---|---|
| | Factors | Items | UDUS (Mean) | BUK (Mean) | ABU (Mean) | U M (Mean) | 1 | 2 | 3 | 4 | 5 |
| 1 | Governance | 14 | 25.14 | 29.25 | 31.94 | 22.82 | 1-14 | 15-28 | 29-57 | 44-57 | 58-70 |
| 2 | General process | 6 | 20.86 | 15.7 | 20.05 | 19.77 | 1-6 | 7-12 | 13-18 | 19-24 | 25-30 |
| | Acquisition & circulation | 11 | 36.57 | 38.06 | 37.21 | 37.73 | 1-11 | 12-22 | 13-33 | 34-44 | 45-55 |
| | Special collection | 4 | 12.86 | 12.83 | 13.21 | 12.83 | 1-4 | 5-8 | 9-12 | 13-16 | 17-20 |
| | Technical processes | 8 | 25.00 | 24.50 | 26.26 | 24.23 | 1-8 | 9-16 | 17-24 | 25-32 | 33-40 |
| 3 | People | 10 | 21.71 | 23.42 | 25.63 | 24.99 | 1-10 | 11-20 | 21-30 | 31-40 | 41-50 |
| 4 | Physical | 12 | 38.14 | 30.25 | 39.95 | 35.45 | 1-12 | 13-24 | 25-36 | 37-48 | 49-60 |
| 5 | Awareness of incidences | 21 | 42.00 | 42.25 | 52.26 | 49.10 | 1-21 | 22-42 | 43-63 | 64-84 | 85-105 |
| | Perceptions & Attitudes | 14 | 46.43 | 46.92 | 48.05 | 46.32 | 1-13 | 14-26 | 27-39 | 40-52 | 53-65 |
| | Awareness of problems | 8 | 14.29 | 16.78 | 19.91 | 23.45 | 1-8 | 9-16 | 17-24 | 25-32 | 33-40 |

Key: 1= Not-Implemented, 2=Planning stage, 3= Partial implementation, 4=Close to completion, and 5=Full implementation
Key: UDUS=Usman Danfodiyo University Sokoto; BUK=Bayero University Kano, ABU=Ahmadu Bello University; UM=University of Maiduguri

For the governance factor, the result indicates that, the library of ABU has the highest mean scores (31.94) followed by BUK library, (29.25), UDUS library, (25.14) and UM library (22.82). The result indicates partial implementation of collection security governance in all the four universities. This indicates that none of the libraries have achieved near or full completion level of implementation and infer an area which the library management should be concerned with. Soete (1999) explained that security management in libraries should involve the formulation of proper governance with a written plan for library security, the designation of a library security officer and the involvement of a team in tackling library security issues. The plans and policies should be continuously assessed and evaluated to ensure security sustainability. McComb and Dean (2004) highlighted the importance of risk assessment in libraries so that the library's assets can be itemized and mapped to past security breaches to enable an effective security policy, processes and plans are in place.

In terms of the implementation of the general operation processes, the libraries of UDUS (20.86), ABU (20.05) and UM (19.77) identified close to completion status, while BUK library is at the partial implementation stage (15.17). The result of the findings show that, all the libraries of the four universities demonstrated close to completion level of implementation security for the acquisition and circulation processes. This is revealed by the mean scores of all the university libraries, which ranged between a mean of 36 and 38. In terms of securing special collections, all university libraries demonstrated close to completion level of implementation. The status of implementation of the technical processes shows that, two of the four university libraries demonstrated close to completion level of implementation, with the mean scores of 26.26 for ABU library and 25.00 for UDUS library, while the rest are at the partial implementation stage. Overall the results show adequate implementation levels even though none has achieved the full implementation stage. Jordan (1999) proposed that to ensure continued access and longevity of collections a proper framework involving the establishment of a repair and binding unit, preservation and duplication unit (for microforms, digital materials and photo duplicating) should be considered.

For people factor, the mean score of all the four university libraries fell within the range of 21-25 which is an indication of partial implementation. Jordan (1999) emphasised on the training of users and staff in a properly planned programme. For example, libraries should consider including security issues and responsibility in user and staff education or orientation programmes.

The physical factors of collection security of the instrument explore both the use of electronic and non-electronic strategies for the management of collection security in university libraries. Three of the four university libraries studied reported close to completion status of implementation (ABU, UDUS and UM), while BUK library indicated partial implementation. Soete (1999) mentioned the importance of controlling and monitoring library premises (exits, entrances, stacks and reading areas). McComb and Dean (2004) proposed that to ensure physical security close attention should be given to the building design, lighting, space planning for shelves, reading areas, office areas, windows, and doors. McComb and Dean also observed that surveillance is lacking in libraries and to add to the problem, keys to libraries' vital places are often kept unsecured or open to easy sight of users and staff.

The security culture factor is made up of three subcomponents of awareness of collection security incidences, perception values and attitudes and awareness of obstacles and problems towards collection protection. The result of the first subcomponent of the awareness of collection security incidences shows that all the four university libraries indicated being aware of collection security incidences in their libraries. This however is contrary to the findings of other studies conducted by Salaam (2004); Ajayi and Omotayo (2004); and Maidabino (2010) which reported staffs' lack of awareness of collection security incidences in university libraries. With regards to the perception, value and attitudes towards collection protection items, all of the university librarians agree with the positive perception, value and attitudes of collection protection. This is indicated by the mean score of all the four university libraries which ranged within 46-48. The last part of security culture factor is the awareness of obstacle and problems on collection security management. The findings of the study revealed that respondents from UM (mean, 23.45) and ABU (19.91) partially agree with the existence awareness of obstacles and problems in the management of collection security, while BUK (mean, 16.78) and UDUS (14.29) on the other hand disagree.

## RELIABILITY ASSESSMENT

The instrument is also subjected to reliability test. The result of the factor analysis shows the commonalities of the data. Zhao (2009) advised researchers to check communality of each items and drop those that have the smallest commonality score. This study adopts Zhao's approach, where any item with communality scores of less than 0.6 is dropped. In this study the following rule of thumb is followed, where an alpha score of 0.9 is considered as excellent, 0.8 good, 0.7 acceptable, 0.6 questionable, 0.5 Poor and less than 0.5 as unacceptable (George and Mallery 2000; Gliem and Gliem 2003).

## Governance Factor

The 14 items scored excellent Cronbach's Alpha scores of over 0.9. The overall governance total score was 0.95 indicating the reliability of the governance construct. The results infer that generally more than a third (44.45%) of respondents disagree or strongly disagree that collection security governance were in place in Nigerian university libraries (Table 5). This indicates that proper governance to secure collections need to be more cohesive and transparent so that both staff and users are aware of their responsibility and its existence.

Table 5: Governance Items, Reliability Value

| Items | Factor 1: Governance | Cronbach's Alpha |
|---|---|---|
| 1.1 | There is a library security management (LSM) team or committee chaired by a senior personnel in the library. | .956 |
| 1.2 | The LSM team comprises staff from various departments of the library & security department of the university. | .953 |
| 1.3 | The LSM team formulates objectives and strategies for library and collection security. | .951 |
| 1.4 | The LSM team has clearly assigned and reporting roles with regard to collection security compliance and improvements. | .953 |
| 1.5 | The LSM team includes personnel from other departments within and outside the library. | .955 |
| 1.6 | The LSM team members have the necessary experience and knowledge about collection security issues and management. | .952 |
| 1.7 | The LSM team have the authority to manage and ensure compliance | .953 |
| 1.8 | Documents about library collection security policies and procedures related to access and use of collections are available. | .952 |
| 1.9 | The security policies and procedures are regularly reviewed and updated | .957 |
| 1.10 | The LSM team undertakes risk assessments and emergency management | .956 |
| 1.11 | The LSM team defines, classifies and determines risks to the various collections in the library | .954 |
| 1.12 | There are proper written and tested procedure for emergencies and breach of security (managers, staff knows what to do) | .955 |
| 1.13 | The LSM receives regular and current reports on security breach incidences | .957 |
| 1.14 | The results of assessments and action plans are communicated to stakeholders | .954 |
| | TOTAL | 0.956 |

| **Items 1.1-1.14: Governance ratings** | 1 | 2 | 3 | 4 | 5 |
|---|---|---|---|---|---|
| Percentage | 27.57% | 16.93% | 22.88% | 13.04% | 19.57% |

1=Strongly Disagree, 2 = Disagree, 3 = Fairly Agree, 4 =Agree, 5= Strongly Agree

## General Operations and Processes

The pilot results show the reliability and internal consistency of each items listed under the general operations and processes component of the instrument. There are a total of 29 items listed under 4 sub-sections comprising 6 items on general operational processes, 11 items on acquisition and circulation processes, 4 items on special collection of the library and 8 items on physical and technical services. All items show good reliability score of above 0.8 and deemed suitable for further analyses (Table 6).

## People Awareness and Training Factors

The ten items listed under people factor of collection security management indicated an acceptable reliability Cronbach's Alpha score of between 0.7 and 0, and on the whole scored a value of 0.811, which suggested good internal consistency and reliability of items in the instrument. The results of respondents' ratings indicate that more than 50% of respondents were not positive about their library's status in implementing training and awareness programmes for staff. Partial implementation is therefore indicated (Table 7). This infers that the four university libraries need to pay more attention to this factor.

Table 6: Operations and Processes Items, Reliability Value

| Items | Factor 2: General Operations and Processes | Cronbach's Alpha |
|---|---|---|
| | **General** | |
| 2.1 | There are written policies, rules and procedures related to collection access, use, controls and protection formulated by the various departments involved in the life cycle of library collections | .847 |
| 2.2 | The assets in the departments are identified and valued | .846 |
| 2.3 | Breach to the collection in every departments are systematically recorded, reported, reviewed and acted upon | .847 |
| 2.4 | Departments undertake backups and recovery of unsecured or lost collections | .849 |
| 2.5 | Staff acknowledge their awareness and acceptance of responsibility for security | .852 |
| 2.6 | Staff are aware of collection security weaknesses and undertake preventive measures and remedial actions | .850 |
| | **Acquisition, circulation** | |
| 2.7 | There are written policies and procedures related to collection selection, acquisition, licensing, subscription, and disposal. | .851 |
| 2.8 | All materials acquired by the library are recorded and numbered | .858 |
| 2.9 | There are written policies and procedures related to the registration, access to the library buildings and the various departments | .851 |
| 2.10 | Regular collection inventory is conducted and reported to ascertain total collection, detect security breach (lost, misplaced, theft, decayed, damaged items) | .856 |
| 2.11 | There is a written manual that details rules about use of collections for both print and electronic collections | .858 |
| 2.12 | Circulation and borrowing of collection are monitored to identify delinquent borrowers as well as shelve space planning | .849 |
| 2..13 | Collection lost, damaged, vandalized are recorded and report prepared and submitted to LSM team | .847 |
| 2.14 | Valuable collections on open stacks are identified, recorded and reported (inventory) | .851 |
| 2.15 | A Collection Maintenance Unit is available to undertake simple rebinding, repairs, maintain stacks and space planning | .855 |
| 2.16 | Rules about duplicating collections and infringement of copyrights are formulated and clearly displayed | .855 |
| 2.17 | There are documented procedures about collection recovery | .857 |
| | **Special collections** | |
| 2.18 | Rules regarding access, use, duplication and protection to special collection is indicated to staff and users (registration, sign log book) | .852 |
| 2.19 | Adequate staff are placed to monitor use of rare collections | .858 |
| 2.20 | The library has a preservation policy which includes duplications, digitization, and microforms | .860 |
| 2.21 | Rare collections are adequately protected from natural disaster (proper storage, fumigation, temperature control | .856 |
| | **Technical and Cataloguing processing** | |
| 2.22 | There are written policies and procedures related to collection description, process time and availability period of all materials acquired by the library | .850 |
| 2.23 | All materials purchased or subscribed are described and documented in the library system to establish ownership | .854 |
| 2.24 | All collections are marked with magnetic strips to establish ownership and to detect unauthorised removal | .859 |
| 2.25 | Unprocessed items are cleared within specified time period to avoid security risks and access to them are controlled (backlogs) | .850 |
| 2.26 | There are written policies and procedures related to access and control to computer terminals and computer systems placed in the library | .850 |
| 2.27 | Backup and recovery policies and procedures are documented - All records of collections are backed up regularly and kept off-site | .847 |
| 2.28 | Access to computer workstations at the library is controlled with password authentication | .862 |
| 2.29 | There are appropriate measures in place that controls and prevent users from installing and using unauthorised software in library workstations | .845 |
| | TOTAL | 0.857 |

| Processes and Operations | 1 | 2 | 3 | 4 | 5 |
|---|---|---|---|---|---|
| Items 2.1-2.6: General processes | 4.92 | 4.64 | 26.78 | 11.48 | 52.19 |
| Items 2.7 – 2.17: Acquisition, circulation | 3.43 | 4.62 | 10.28 | 11.03 | 70.64 |
| Items 2.18 – 2.21: Special collections | 5.51 | 4.72 | 14.57 | 19.69 | 55.51 |
| Items 2.22 – 2.29: Technical services | 8.20 | 3.69 | 21.93 | 11.68 | 54.51 |

1=Strongly Disagree, 2 = Disagree, 3 = Fairly Agree, 4 =Agree, 5= Strongly Agree

Table 7: People Issues of Collection Security, Reliability Value

| Items | Factor 3: People Issues | Cronbach's Alpha |
|---|---|---|
| 3.1 | Library staff at various level are aware of the contents of the library's collection security policy, procedures and rules. | .792 |
| 3.2 | Manuals and leaflets about the library's collection security regulations, policy and procedures are available to staffs. | .827 |
| 3.3 | Staff and users are informed about the importance of collection security and informed to report security breach incidences. | .800 |
| 3.4 | Signage both print and electronic informing about collection security and protection are easily seen by both staff and users. | .771 |
| 3.5 | There is a written document which indicates the required competency and learning outcomes on collection security management and processes | .785 |
| 3.6 | Staff are trained to monitor and handle collection security breaches on their own | .792 |
| 3.7 | There are organised activities and training being offered to increase staff and user awareness | .798 |
| 3.8 | Staff and users are given training to understand the content of the collection security policy and how to handle security breach incidences | .781 |
| 3.9 | Staff are trained to record and report all breach of collection security | .768 |
| 3.10 | Staff are trained to handle delinquent borrowers and users | .818 |
| | TOTAL | 0.811 |

| Factor Items | 1 | 2 | 3 | 4 | 5 |
|---|---|---|---|---|---|
| 3.1-3.10: People Issues | 11.48 | 12.79 | 33.93 | 12.13 | 29.67 |

1=Strongly Disagree, 2 = Disagree, 3 = Fairly agree, 4 =Agree, 5= Strongly Agree

## The Physical Factors of Collection Security

All 12 items listed under physical factors shows overall acceptable reliability scores based on the Cronbach's Alpha score of above 0.7 (Table 4). Students were asked to rate on their agreement on the perceived status of physical and technological practices on collection security management in their libraries. The ratings on the ten items show that over 60% of respondents agree and strongly agree on full implementation of physical factors related to collection security in their libraries (Table 8).

Table 8: People Issues of Collection Security, Reliability Value

| Items | Factor 4: Physical Factors (Non-electronic and Electronic) | Cronbach's Alpha |
|---|---|---|
| 4.1 | Security systems are placed at entrance, exit and stack areas in the library to prevent unauthorised removal of collections (Electronic anti theft system, visual cameras, smoke detection system, CCTV, magnetic detection system) | .776 |
| 4.2 | Student and staff ID is required at entrances to general and special collections area | .718 |
| 4.3 | Preventive measures are put in place in libraries to protect collections (restricting staff to control areas, computer laboratories, rare collections area, grills for windows) | .708 |
| 4.4 | Staff and security personnel have scheduled patrols within building parameters in open stacks and rare collections areas. | .709 |
| 4.5 | Fire alarms are placed in strategic areas of the library and these equipments are maintained and tested (through fire drills) | .721 |
| 4.6 | Rules and procedures to access restricted areas during and after opening hours are clearly documented and understood by all staff | .716 |
| 4.7 | All OPAC stations, PC/Internet workstations are protected from unauthorised access (through passwords and user IDs) | .682 |
| 4.8 | Maintenance of all OPAC and Internet workstations are scheduled regularly | .705 |
| 4.9 | Periodic random checks are carried out on users and staff who enters or exit the library is carried out (checking IDs and personal belongings) | .746 |
| 4.10 | Collection security is given consideration when planning the layout of shelves, sitting and reading areas, placement of fire prevention equipment. | .713 |
| 4.11 | Password requirements are in place to access online databases, library systems and electronic resources | .734 |
| 4.12 | Firewalls and intrusion detection systems are installed to hinder unauthorised user access to library systems and databases | .767 |
| | TOTAL | 0.743 |

| **Factor Items** | **1** | **2** | **3** | **4** | **5** |
|---|---|---|---|---|---|
| **4.1-4.12: People issues** | 6.01 | 10.79 | 20.90 | 11.20 | 51.09 |

1=Strongly Disagree, 2 = Disagree, 3 = Fairly agree, 4 =Agree , 5= Strongly Agree

## Security Culture in Libraries

When tested for reliability, the Security Culture items under each components shows an overall score of 0.815 based on Cronbach's Alpha coefficient which makes this construct acceptable for further analysis as described by Garson (1998) and Suhazimah (2007) (Table 9). The results indicate that staff in the sample libraries have acceptable degree of awareness about security breach incidences and have a positive perception and attitudes towards collection protection.

Table 9: Security and Collection Security, Reliability Value

| Items | Factor 5: Security Culture | Cronbach's Alpha |
|---|---|---|
| | **Aware of these incidences in the library** | |
| 5.1 | Books not returned by borrowers | .815 |
| 5.2 | Pages torn from books | .791 |
| 5.3 | Articles torn from journals | .791 |
| 5.4 | Book borders and pages are doodled | .832 |
| 5.5 | Book spine ripped off to remove magnetic strip | .783 |
| 5.6 | Pages cut out from whole text with binding intact | .791 |
| 5.7 | Book spine are torn due to wrong handling or pulling at the spine | .803 |
| 5.8 | Books are wrongly shelved deliberately | .783 |
| 5.9 | Pages are eaten by book insects or silver fish | .782 |
| 5.10 | Pages are stained and yellow | .846 |
| 5.11 | Rare items are reported lost | .782 |
| 5.12 | Books reported stolen | .777 |
| 5.13 | Controlled items such as thesis are found being photocopied without permission | .832 |
| 5.14 | Brittle pages of old books and manuscripts | .827 |
| 5.15 | Collection ruined because of natural disaster (flooding) | .822 |
| 5.16 | Rare items are found on open shelves | .805 |
| 5.17 | User caught borrowing using someone else's ID | .786 |
| 5.18 | Rare books of value are not insured | .822 |
| 5.19 | Excessive downloads from subscribed online databases | .783 |
| 5.20 | Staff keeps books processed or otherwise for personal use | .783 |
| 5.21 | Staff allows friends, family members to borrow restricted item | .793 |
| | **Perceptions, value and attitudes towards collection protection** | |
| 5.22 | Library management is committed and appreciates the importance of collection security | .818 |
| 5.23 | Staff at knows their individual library security responsibility | .806 |
| 5.24 | Library staff knows how to handle any security incidences | .841 |
| 5.25 | Library staff at all levels comply with security rules and procedures of the library | .827 |
| 5.26 | Protection of collections is important so that the library can deliver its services effectively | .816 |
| 5.27 | Protection of collections is important to support the library's teaching, learning and research objectives | .816 |
| 5.28 | The success of the library service depends on the availability of its collections when it is needed | .816 |
| 5.29 | The success of the library service depends on assurance that the integrity of the collections are maintained (contents are kept intact) | .816 |
| 5.30 | The success of the library service depends on ensuring the confidentiality of the collection by users (duplication of materials are controlled to protect ownership rights) | .814 |
| 5.31 | My library regularly undertake inventory of the collection to ascertain loss, damaged or vandalism | .816 |
| 5.32 | My library will prioritize and act upon action plans based on inventory reports | .838 |
| 5.33 | My library considers breach of security incidences a serious issue and initiates preventive recovery plans. | .830 |
| 5.34 | The status of library security and collection security are communicated to staff and stakeholders | .616 |
| | **Awareness of obstacles and Problems** | |
| 5.35 | My library faces budget constraints to implement CSM | .816 |
| 5.36 | Management in my library are not committed to protecting the library collection | .835 |
| 5.37 | Management in my library are not aware of the extent of security incidences occurring in the library | .816 |
| 5.38 | Management and staff in the library is hiding the true situations about collection violation incidences | .816 |
| 5.39 | My library do not have a standard documented implementation plans concerning collection security | .816 |
| 5.40 | The staff in library at all levels are unskilled on how to resolve collection security issues | .816 |
| 5.41 | My library finds it difficult to justify the value of a collection security programme | .816 |
| 5.42 | Users know how to outsmart the library security system | .816 |
| | TOTAL | 0.815 |

| Factor Items | 1 | 2 | 3 | 4 | 5 |
|---|---|---|---|---|---|
| **Items 5.1-5.21: Awareness of common security incidences** | 10.6 | 12.8 | 32.8 | 32.8 | 11.0 |
| **Items 5.22 – 5.35: Perception, Value and attitudes towards collection protection** | 2.93 | 00 | 8.20 | 18.03 | 70.84 |
| **Items 5.36 – 5.43: Awareness of obstacles and problems** | 12.09 | 35.66 | 14.75 | 22.95 | 14.55 |

1=Strongly Disagree, 2 = Disagree, 3 = Fairly agree, 4 =Agree , 5= Strongly Agree

## CONCLUSION

In this study, reliability refers to the degree to which the scales properly assess collection security-related implementations in university libraries. To establish the reliability of the instrument, the researcher first collated and compiled collection security issues derived from published literature and used the items to design the assessment instrument. The result indicates that, overall, all the items covered by the five factors in the proposed instrument were found to have an acceptable measure of reliability as the majority of the items registered Cronbach's Alpha scores between 0.7 and above 0.9 . Garson (1998) justified the sufficiency of these Alpha scores when he opined that, "In the early stages of research, reliability in the range of .05 to 0.6 is sufficient". Also, Nunnally (1978) and Bartlett et al. (2001) proposed that "... in general, an alpha level of 0.5 is acceptable". However, Garson (1998) proposed that acceptable alpha level in an exploratory research should be at least 0.7 or higher to retain an item in an 'adequate' scale. The pilot study indicates that the assessment instrument is a reliable measure to assess the level of implementation of collection security implementation in university libraries and the implementations of collection security management are at different stages in the Nigerian University libraries.

The instrument indicates the following.

- Governance related to collection security management is only partially implemented at the ABU and BUK libraries. The two universities are however much older universities, which could be a reason for the implementation of governance aspect of collection security. While, the other university libraries reported non-implementation of the governance aspect of collection security. Boss (1994) amongst other issues identified that poor policies and procedures, lack of security plans and manuals are triggers for security breaches. Governance should be properly established as university libraries grow in size as Olorunsola (1987) discovered a significant relationship between high rates of security problems and the growth of universities in Nigeria.
- Mean ratings from all the university librarians reported close to completion in the implementation of the general processes, and those that support CSM in acquisitions, circulation, special collections and technical processes. These are issues related to written policies, regulations and rules regarding access, use, control, protection of collections and procedures for backups and recovery are in place. The majority of respondents perceived the implementation status to be in place and close to completion.
- In terms of making staff trained and aware, the mean ratings by all librarians at the four universities reported partial implementation. The librarians are aware of training programmes. About a third (>30%,>18) perceived that signage and proper programmes should be clear and widely disseminated to staff so that they can assimilate issues related to collection security policy and know how to handle security incidences, indicating areas that the library security team should focus upon.
- With regards to the physical aspect of security, 50% (31) of the university librarians reported close to completion status while 50% (31) also indicated that their libraries are at the partial stage of implementation. Even though some of the libraries that perceived that the physical mechanism of protecting library collections are in place there are still doubts about their library's strength in terms of installing electronic devices to safeguard collections. This is expected as the installation of such devices are expensive and beyond the means of most academic libraries in Nigeria. The results infer that a more comprehensive strategy needs to be implemented in securing library premises and the collection effectively.

- Collection security culture is partially in existence but needs improvement. Almost all the libraries have very positive attitudes and values towards collection protection but only 50% (31) are aware of the problematic situations that prevent the smooth implementation of a collection security programme. Many are not aware that excessive downloads from subscribed databases is a security breach (65.5%, 40) and that rare books of value needed to be insured (52.4%, 32).

The American Library Association in a published document about library security (Library Security 2001) encapsulated most of the five factors in the house model. ALA had emphasised the need to protect library buildings, their employees and users, suggesting preventing actions to combat collection loss, formulating disaster plan and security policy, assigning and training staff to handle security issues. The Association of College and Research Libraries (2003, 2006) has also published two guidelines for handling theft in libraries and for handling rare and special collections. However, studies that provide an instrument that can be used to assess collection security implementation cannot be located. In this paper we propose to close this gap by introducing an assessment instrument that has been found to be reliable and usable in assessing the collection security management in libraries in a more holistic approach. The robustness of this instrument need to be further tested and this will carried out in the second phase of this study.

## REFERENCES

Abifarin, A. 1997. Library stock security: The experience of the University of Agriculture Abeokuta, Nigeria. *Library and Archival Security*, Vol.14, no.1: 11-19.

Ajayi, N.A & Omotayo, B.O. 2004. Mutilation and theft of library materials: Perception and reactions of Nigerian students. *Information Development*, Vol.20, no. 1: 61-66

Ajegbomogun, F.O. 2004. Users' assessment of library security: A Nigerian university case study. *Library Management*, Vol.25, no.8/9: 386-390.

Alao, I.A., Folorunso, A.L and Saka, H.T. 2007. Book availability in the University of Ilorin College of Health Sciences Library. *World Libraries*, Vol.17, no.2. Also available at http://www.worlib.org/vol17no2/alao_v17n2.shtml.

Allen, J. and Westby, J.R. 2007. Characteristics of effective security governance. *Governing for Enterprise Security (GES) Implementation Guide* (CMU/SEI-2007-TN-020). Software Engineering.

Allen, S.M. 1997. Preventing theft in academic libraries and special collections, *Library Archives & Security,* Vol.14, no.1: 29-43.

Ameen, K. & Haider, S.J. 2007. Evolving paradigm and challenges of collection management in university libraries of Pakistan, *Collection Building,* Vol.26, no.2: 54-58.

Ardndt, D.A. 1997. Problem patrons and library security, *Legal Reference Services Quarterly*, Vol.19, no.1 & 2: 19-40.

Association of College & Research Libraries. 2003. *Guidelines regarding thefts in libraries*. Available at http://www.ala.org/ala/mgrps/divs/acrl/standards/guidelines regarding thefts.cfm.

Association of College & Research Libraries. RBMS Security Committee. 2006. *Guidelines for the security of rare books, manuscripts and other special collections*. *C & RL News*, Jul/Aug: 426-433. Available at http://www.ala.org/ala/mgrps/divs/acrl/standards/securityrarebooks.cfm.

Atkins, S.S. and Weible, C.L. 2003. Needles in a haystack: Using interlibrary loan data to identify materials missing from a library's collection. *Library Collections, Acquisitions, & Technical Services*, Vol.27: 187-202.

Aziagba, P.C. and Edet, G.T. 2008. Disaster control planning for academic libraries in West Africa. *The Journal of Academic Librarianship*, Vol.34, no.3: 265-268.

Bartlett, J.E., Kotrlik, J.W.and Higgins, C.C. 2001. Organizational research: Determining appropriate sample size in survey research. *Information Technology Learning and Performance Journal.* Vol.19, no.1: 43-50.

Bello, M.A. 1998. Library security, material theft and mutilation in technological university libraries in Nigeria. *Library Management*, Vol.19, no.6: 379-383.

Boss, R. 1984. Collection security. *Library Trends*, Vol.18, no.1: 39-48.

Brown, K.E. and Patkus, B.L. 2007. *Collection security: Planning and prevention for libraries and archives*. Northeast Document Conservation Centre. Available at: http://www.nedcc.org/resources/leaflets/3Emergencymanagement/11collectionsSecurity.php.

Corporate governance task force report. 2004. Information security governance: a call for action. Available at http://cyberpartnership.org/infosecGov4 04.pdf.

Cowan, J. 2003. Risk management, records and gaming report. *Clinical Governance: an International Journal*, Vol.8, no.3: 275-277.

Evans, G.E., Amodeo, A.J. and Carter, T.L. 1999. *Introduction to Library Public Services*. 6th ed. Libraries Unlimited, Inc. Englewood: Colorado.

Ewing, D. 1994. Library security in the UK: are our libraries of today used or abused? *Library Management*, Vol.15, no.2: 18-26.

Garson, G.D. 1998. Reliability analysis. Available at http://faculty.chass.ncsu.edu /garson/PA765/reliab.htm.

George, D. and Mallery. 2000. *SPSS for windows step by step: a simple guide and reference*. 3rd edition. Allyn & Bacon.

Gliem, A.J. and Gliem R.R. 2003. Calculating, interpreting, and reporting Cronbach's Alpha reliability coefficient for likert-type scales. Midwest Research to Practice Conference in Adult, Continuing, and Community Education: The Ohio State University, Columbus, October 8-10, 2003, 82-88.

Hill, D. A. and Rockley, L. E. 1981. *Security: its management and control*. Business Book: London.

Holt, G.E. 2007. Theft by library staff. *The bottom line: managing library finances*, Vol.20, no.2: 85-92.

Houlgate, J. and Chaney, M. 1992. Planning and management of a crime prevention strategy, in: *Security and Crime Prevention in Libraries* Edited by Michael Channey and Alan F. MacDougall. Gower Publishing, Gower House: 46-49. ISBN 1857420144.

Lorenzen, M. 1996. Security issues of academic libraries. *ERIC Document. No. ED396765.*

Lowry, C.B. and Goetsch, L. 2001. Creating a culture of security in the university of Maryland libraries. p*ortal: Libraries and Academy*, Vol.1, no.4: 455-464.

Luurtsema, D. 1997. Dealing with book loss in an academic library. *Library Archives & Security*, Vol.14, no.1: 21-27.

Maidabino, A. A. 2010. Collection security issues in Malaysian academic libraries: An Exploratory Survey. *Library Philosophy and Practices*, Available at http://unllib.unl.edu/LPP/lpp2010.htm.

Momodu, M.A. 2002. Delinquent readership in selected urban libraries in Nigeria. *Library Review.* Vol.51, no.9: 469-473.

Nunnally, J.C. 1998. *Psychometric theory*. 2nd. Ed. New York: McGraw-Hill.

Omoniyi, J.O. 2001. The security of computer and other electronic installations in Nigerian university libraries. *Library Management*, Vol.22, no.6/7: 272-277.

Purtell, T. 2007. A new view on IT risk. *Risk Management,* Vol.54, no.10: 28.

Saffady, W. 2005. Risk analysis and control: Vital to records protection. *Information Management Journal,* Vol.39, no.5: 62-64.

Rude R. and Robert, H. 1993. Theft, dissemination and trespass: Some observations on security. *Library and Archival Security*. Vol. 12, no.1: 18-21.

Salaam, M.O. 2004. The treatment of other libraries' books by Nigerian university libraries. *Library & Archival Security*, Vol. 19, no.1: 47-51

Shuman, A.B. 1999. *Library security and safety handbook: Prevention, policies and procedures*. Chicago : American Library Association.

Suhazimah, D. 2007. The antecedents of information security maturity in Malaysian public services organizations. Ph.D. thesis, Faculty of Business and Administration, University of Malaya, Malaysia.

Swartzburg, G.S., Bussey, H. and Garretson, F. 1991. *Libraries and archives: Design and renovation with a preservation perspective.* In *Safety, Security, Emergency Planning, and Insurance*. London: Scarecrow Press: 147-169.

Wu, Y.D. and Liu, M. 2001. Content management and the future of academic libraries. *The Electronic Library,* Vol.19, no.6:432-439.

Verton, D. 2000. Companies aim to build security awareness. *Computerworld,* Vol.34, no.48: 24.

Zhao, N. 2009. The minimum sample size in factor analysis. Available at, http://www.encorewiki.org/display/nzhao powered by Atlassian Confluence, the Enterprise Wiki.